\documentstyle[psfig,11pt]{article}
\newcommand{\be}{\begin{equation}}
\newcommand{\ee}{\end{equation}}

\begin{document}

\begin{center}
\large{\bf WEST-SIDE STORY}\\
\large{\bf (On the History of Density-Wave Spiral Theories in the 1960s)}\\
\vspace{0.25in}
Leonid Marochnik\\
Space Telescope Science Institute, Retired\\
lmarochnik@comcast.net\\
\end{center}

\begin{abstract}

This paper deals with the history of Density-Wave Spiral Theories in the 1960s. 
The motivation to write the paper was the publication of two papers on the 
history of these theories (Pasha 2004a, b). Pasha's papers tell only a part 
of the story that took place on the Western side of the Iron Curtain in the 1960s. 
But giving only a part of the full story is a distortion of historical truth. 
Important work done on the Eastern side of the Iron Curtain is still little known in the West. 
In this paper, I fill the gaps and correct chronological inaccuracies in Pasha's story and 
mention facts that are still unknown (or little known) to the astronomical community in the West. 
I also give my recollection of the development of Density-Wave Spiral Theories in the 1960s.  
\end{abstract}

\section{ DENSITY-WAVE THEORIES IN THE 1960's (MY RECOLLECTION)}

The interesting article by I.Pasha discusses the early history of the Density Wave Theory in the 1960's. 
This was a period of rapid development when the foundations of the theory were laid. 
Pasha's articles are of interest particularly because in addition to a fascinating description of 
the development and conflicts of ideas, they are based on a great number of private communications 
with the developers of this theory such as A. Kalnajs, C.C. Lin, D. Lynden-Bell, F. Shu, A. Toomre, and others. 
These people made fundamental contributions to the development of the theory and laid its foundation. 
I was also a participant in the development of this theory in the 1960's. 
In this paper, I want to share my recollections, fill in gaps in the history written by Pasha 
and address some historical inaccuracies. As I will show in this article, the real history of the 
Density Wave Spiral Theory in the 1960's frequently was not quite as described by Pasha 
and occasionally quite different.
\\

The history described by Pasha has at least one fundamental flaw. 
It is based exclusively on contributions and recollections of people who 
worked on the West side of the Iron Curtain. 
\\

Actually, I was surprised that the Russian astronomer/historian I.Pasha completely 
ignored all the work done on the Eastern side of the Iron Curtain in the 1960s. 
As a matter of fact, being a Russian speaker he had a unique opportunity to read papers 
that have never been translated into English and to present to the world science 
community important contributions still unknown in the West. 
\\

Thus, I would like to describe the efforts in the 1960's of a number of scientists 
behind the Iron Curtain who have contributed in no small measure to the development 
of the Density-Wave Spiral Theories. In fact, the work of these scientists challenged some 
of the basic ideas and results of their western counterparts. 
The more complete history of the development of the field can be found in the book 
"The Milky Way Galaxy" (Gordon \& Breach) by Marochnik and Suchkov (1996).
\\

My graduate students, associates and I were the first to begin work on the Density Wave Theory 
in the former Soviet Union. Therefore, for this particular period of time (1964-1969) 
I have to speak about our own work. 
\\

In reading Ogorodnikov's book (1958) "Stellar Dynamics" in 1963 I immediately saw 
analogies and differences between colissionless plasma and encounterless stellar systems (e.g., galaxies). 
One of the main similarities was the straightforward analogy between the Lorentz force 
acting on an electron in a magnetic field and the Coriolis force acting on a particle in 
rotating gravitational systems. And the other analogy was the possibility of collective 
phenomena such as waves in both plasma and gravitation. 
The main difference between the two systems was that the forces acting between plasma 
particles were repulsive, while the forces acting on stars were attractive.
\\

Using the first analogy, I showed that under some conditions collisionless 
(or better to say, encounterless) rotating stellar systems can be described by equations 
of the magneto-hydrodynamics type (Marochnik, 1964). I did it in the same way as it was 
done by Chew et al. (1956) and independently by Rudakov and Sagdeev (1958) for 
magnetized collisionless plasma. Generalizations to the case of differential rotation 
are due to Kato (1968) and Hunter (1979). 
A generalization to the case of a differentially rotating infinitely thin disc is due to Berman and Mark (1977).  
\\

Using my new hydrodynamic equations, I found (Marochnik, 1966) that density waves of collective nature 
could exist in rotating stellar systems. These were waves of the same nature that were found by 
Lynden-Bell (1962) who used a much more complicated method of the kinetic Boltzmann equation. 
This idea is still in use in current numerical works on spiral density waves (Korchagin, 2004). 
\\

The possibility of the existence of density waves in gravitational systems was a non-trivial 
question at that time. In the non-magnetized plasma, repulsive Coulomb forces between electrons 
provided the necessary "elasticity" that led to wave propagation. 
In magnetized plasma, besides "elasticity" due to Coulomb forces, there exists "elasticity" 
due to the magnetic field. There was no "elasticity" due to Newton forces in gravitation. 
However, there was "elasticity" due to the Coriolis forces. 
Therefore, one would expect to find density waves in a gravitational system due to its rotation.
\\

It was probably Sweet (1963) who was the first to advance the idea of collective processes 
in gravitational systems. The title of his paper was "Cooperative Phenomena in Stellar Dynamics", 
and he applied directly methods developed for plasma physics. 
\\

It so happened that I did not pay any attention to Lin and Shu (1964, 1966) papers until the year 1967. 
This was probably because of my purely academic interest in the nature of density waves in 
gravitational systems in general (Cartesian geometry). At that time collective phenomena in 
gravitation were of great interest to Russian physicists. In particular, Muzafar Maksumov and I 
obtained a generalization of Jeans' critical length (for a gas) for a colissionless gravitational 
system with an arbitrary velocity distribution function. Considering it as an important work, 
Yacov Zeldovich presented it to the most prestigious journal of Russian Academy of Sciences "Doklady" 
(Maksumov and Marochnik, 1965).   
\\

However, the last of our papers from this series of "academic interest only papers" 
(Marochnik and Ptitsyna, 1968)  already contained  an important effect, discovered later 
in the theory of spiral density waves in stellar discs, which became known as an ultra 
harmonic resonance phenomenon. 
\\

Bernstein (1958) discovered this effect for plasma waves propagating in the plane 
normal to the direction of the magnetic field (so-called Bernstein modes). 
He found that due to the resonance between wave frequency and overtones of electron gyro-frequency, 
there exist zones forbidden for wave propagation. Because of analogy between Lorentz and Coriolis 
forces we expected to find it in rotational gravitational systems as well. 
And we found it (Marochnik and Ptitsyna, 1968). 
The physics of this phenomenon is the same in any rotational stellar system independently 
whether or not it is a Cartesian or a curvilinear coordinate system. 
The difference would be only quantitative but not qualitative. 
It is a resonance between wave frequency and overtones of particle's (star's) epicycle frequency 
in the curvilinear (e.g., cylindrical) coordinate frame. 
In rotating Cartesians coordinates, it is a resonance between wave frequency and overtones 
of particle rotating along a Coriolis circle. 
\\

Lynden-Bell (1962) knew of Bernstein's paper and used his mathematical approach to obtain 
the dispersion relation for stellar waves in the Cartesian reference frame. Lin, Yuan and Shu (1969) 
also mentioned the similarity of alternative form of the dispersion equation for density waves 
(equation (1) below) to that of Bernstein (1958) paper. However, Marochnik and Ptitsyna (1968) 
were the first who showed that density waves in rotational Cartesians gravitational systems have "gaps" 
in their spectrum due to resonance between wave and overtones of particles rotating along a Coriolis circle. 
\\

St.Petersburg's physicist Lev Gurevich was probably the first person who called my attention to Lin and Shu work. 
In the Acknowledgment to the Marochnik and Ptitsyna (1968) paper one can find the following. 
"The authors would like to thank L. E. Gurevich for calling our attention to the possible 
relationship between Lin and Shu's spiral waves and the stellar waves discovered in the present investigation". 
Of course, they were waves of the same physical nature.   
\\

We applied the physics of Bernstein modes to spiral density waves in gravitational discs 
and showed the existence of forbidden zones ("gaps") in the wave spectrum due to the resonance 
of a wave with overtones of epicyclical frequency (Marochnik and Suchkov (1968b, 1969a)).  
\\

Later, in the 1970s this effect was re-discovered in the West. Instead of the name that we gave it 
"resonance on overtones of epicyclical frequency", it was named "ultra harmonic resonance" 
in the papers of Lin's camp. It was probably, Contopouls (1970) who first investigated the 
"ultra harmonic resonance" in terms of stellar orbits.  Extensive research of ultra harmonic 
resonance continued for years after that. Unfortunately, I never saw any reference to our original paper.  
\\

Let us go back to the theory of spiral density waves. By the middle of 1967 
I already knew about the work of Lin and Shu (1964, 1966). However, the most important thing 
that pushed me to pay attention to their theory was my talk with Solomon Pikelner sometime in 1967. 
Pikelner was one of the greatest Russian astronomers of that time. 
During the 1960s he also worked on the problem of spiral structure of galaxies. 
However, he believed that spiral structures are gaseous arms with magnetic field, which governs the situation.
\\

During our talk at his home over a cup of tea, he told me that he understood that he was wrong 
(with the magnetic theory) but Lin and Shu were probably, right. It was a turning point for me to 
begin work on spiral density waves. I took up the spiral structure problem in the middle of 1967. 
After Lin and Shu (1966), I began to look for an instability that could maintain spiral density waves 
for a long time. I believed that this was the two-stream instability due to gravitational interaction 
between the flat fast rotating subsystems and "non-rotating" (weakly rotating) subsystems in galaxies. 
So, I invited my then graduate student Anatoly Suchkov to join the research. By the end of 1968, 
a series of four preprints (Marochnik and Suchkov, 1968a, b, c, and d) was published with full 
consideration of the problem. These preprints were in Russian but they had English abstracts and titles. 
The fifth preprint (Marochnik and Suchkov, 1968e) was the English digest of the theory that was given 
in the first four preprints.   
\\

This "spiral series" of our work were well known and accepted by Russian astronomers and physicists 
in 1968-1970 as a further development of the work of Lin and Shu (1964, 1966) and in some aspects as 
its alternative (see below).  Several talks on the theory were presented in the most prominent 
Moscow seminars of those years. These were Vitaly Ginzburg's seminar in the Lebedev Physical Institute, 
Yacov Zeldovich's seminar at the Institute of Applied Mathematics and Landau seminar at 
the Institute of Physical Problems.  There also were many private discussions with Roald Sagdeev 
(a rising star of plasma physics in those days), Solomon Pikelner and Yacov Zeldovich. 
These names can be found in the Acknowledgement to this series of papers. 
These people considered the "spiral series" as a new physical approach to the old astronomical problem (see below).  
\\

I had an invitation to attend the famous "Basel Symposium" of 1969 on spiral structure of galaxies. 
However, the KGB did not allow me to go abroad. So, I had no chance to present to the world's 
astronomical community our work that was ahead of what Lin, Yuan and Shu (1969) did by that time (see below). 
\\

Looking back, I think that the two-stream instability is not as effective in the Milky Way as we 
thought at that time because of a thin fraction of badge-halo stars in the galactic plane. 
But it still applies to counter-rotating discs (V. Korchagin, private communication). 
\\

As I already mentioned, by the end of 1968- beginning of 1969, our "spiral series" of papers have 
already been published (Marochnik and Suchkov, 1968, b, c, d, e; 1969a, b, c).  
At this point, we considered a multi-subsystem model of galaxy, consisting of rotating 
and non-rotating stellar subsystems contrary to Lin's camp who worked with only one stellar subsystem (example 4). 
We showed that the famous Lin, Yuan and Shu (1969) paper employed an incorrect galactic model to compare the 
theory with observations (example 5). We also predicted the existence of discrete spiral modes in galactic discs. 
Lin's camp came to this point several years later (example 3). 
I never saw any references to these results in papers of C.C.Lin, C.Yuan, F.Shu, Y.Y.Lau, 
G.Bertin and other members of Lin's camp.  
\\

We were also the first who studied non-linear interaction of unstable spiral density waves with stars 
of both rotating and non-rotating subsystems. It is important to stress that our non-linear 
treatment was developed for unstable spiral waves of arbitrary type (not necessarily based on two-stream instability).  
\\

Using powerful mathematical methods of plasma physics, we showed that this non-linear interaction 
between waves and stars led to an appearance of a "collision term" on the right hand side 
of the originally "collisionless" Boltzmann equation, which describes star-wave "collisions". 
An immediate consequence was the exchange of angular momentum and energy between spiral density 
waves and stars of both subsystems. Lin's camp came to this point several years later (example 4).  
\\

As I already mentioned, this series of papers (Marochnik and Suchkov, 1968 a, b, c, d; 1969 a, b, c) 
applied physical ideas and mathematical methods of plasma physics that was in a period of rapid development. 
Lin became familiar with plasma physics later in 1970s when Y.Y. Lau joined his team (footnote 78 from Pasha, 2004a).  
I believe we used this advantage in full measure to get ahead in this science competition.
\\

Unfortunately, we were unable to publish the full version of both our linear theory and non-linear 
theory of wave-stars interaction in Astrronomicheskii Zhurnal (Soviet Astronomy) in 1969. 
I remember that the Chief Editor (Evald Mustel) told me that he had no space to publish a work of 45-50 journal pages. 
Astronomicheskii Zhurnal published six issues per year, and the wait for publication at that time was 1-2 years. 
So, we cut our full version (Marochnik and Suchkov 1968 a, b, c, d) by about a factor of two for the 
publication in this journal. The Chief Editor made an exception for our work (it was considered by the 
publication board as extremely timely). Both papers were accepted on November 25, 1968 and appeared in 
print in March-April (No 2) and May-June (No 3) issues, respectively. The time delay was six months only. 
Simultaneously the English version of our theory was published in Astrophysics and Space Science 
(Marochnik and Suchkov, 1969c). This version of our theory was cut by a factor of four.     
\\

Therefore, the short versions of both our linear and non-linear theories (Marochnik and Suchkov, 1969 a, b) 
contained no mathematical details (only final equations). This may have been the reason why these 
papers were not mentioned by scientists in the West. In the end, a full version of the theory 
(four preprints of 1968) was published later in Russian as a regular publication (Marochnik and Suchkov, 1971). 
\\

Below are examples to which I referred above.
\\

\section { ALTERNATIVE FORM OF THE DISPERSION RELATION}

Pasha (2004a)
dedicates several pages of his paper to the history of the so-called alternative form 
of dispersion relation for density waves (1), which he attributes to Lin, Yuan and Shu (1969) 

\be
   1 = { 2 \pi G \mu |k| \over \kappa^2} {2 e^{-x^2} \over x^2} \sum^{\infty}_{s=1}
   {I_s(x^2) s^2 \over s^2 - \big ({\omega - m\Omega \over \kappa} \big)^2 }
\ee

where $\kappa$ is epicyclical frequency, $x = k\sigma/\kappa$ (the rest of notation is below of the eq.(3))
\\

However, the actual chronological sequence of events in the history of equation (1) is the following.
\\

Equation (1) appeared for the first time in the Marochnik and Suchkov's (1968b)
preprint (equation II.19)) and in Marochnik and Suchkov (1969a)
as a journal paper (equation (15))\footnote{For the special case $m=0$ the equation (1) 
has been obtained in Kalnajs' PhD Thesis in 1965.
Of cause, living behind the Iron Curtain in these years, we were unable to know about this unpublished work.
But still, it was a case, which had no direct connection to spiral structure theories because $m$ is
a number of spiral arms, and $m=0$ meant ring-like structures.}.
This preprint was sent to Lin and Shu at the end of 1968
(together with the other preprints, 1968a, c, d, and e)\footnote{I. Pasha referenced often to, 
formally speaking, unpublished but documented
materials like Kalnajs' PhD thesis and Lin's preprint. I believe this is a right
approach when the history is considered. So, I am going to operate in the same terms,
using reference not only to journal papers but to preprints also.}.
Lin, Yuan and Shu (1969) paper appeared in the March 1969 issue of ApJ. 
The very last equation in the Appendix C of this paper is an equation similar to (1). 
However, it is incorrect because it contains the wrong factor $(-1)^s$. 
This issue of ApJ (March 1969) had a glued ERRATA (in front of its back cover), 
which contained the correct equation (1) from Marochnik and Suchkov (1968b). 
\\

The regular journal papers of Marochnik and Suchkov (1969a) and Lin, Yuan and Shu (1969) 
appeared simultaneously in the March issues of ApJ and Astronomicheskii Zhurnal (Soviet Astronomy). 
For some reason, equation (1) is attributed exclusively to Lin, Yuan and Shu.  
\\

The correct form of equation (1) can be also found in the Lin and Shu (1971) paper. 
This is a lecture at the Brandeis University Summer Institute in Theoretical Physics of 1968. 
However, I believe we have to consider the date of publication (1971) rather than the date of lecture alone. 
Remember that the   wrong factor has been fixed in 1969 only. 
\\

\section{WKB}

It is generally accepted now that Lin and Shu (1964, 1966) employed WKB approximation 
(or WKB-type approximation) to obtain the dispersion relation for spiral density waves. 
Pasha as a historian accepts this as an a priori fact. However, the actual history of 
this approach was quite different. 
\\

As a matter of fact, Lin and Shu (1964, 1966) employed the so-called asymptotic approximation 
or small pitch angles approximation. Marochnik and Suchkov (1968a) were the first to show that 
this small pitch angles approximation corresponds to the geometrical optics approximation, 
known in quantum mechanics as the WKB approximation. Lin and Shu (1964, 1966) themselves did 
not mention this important fact.  
\\

It was not just a difference in terminology because of the following. 
\\

Discovery of this fact led us to a very important conclusion that we pointed out 
(Marochnik and Suchkov,1968a, 1969a, c). If there are turning points $R_1$ and $R_2$ in the disk, 
then the WKB approximation yields the "quantization" conditions

\be
  \int_{R_1}^{R_2} k(R,\omega)dR = (n+1/2)\pi
\ee

where $n$ is an integer. This is Equation (I.17) from Marochnik and Suchkov (1968a) and Equation (14) from (1969a).
\\

From this it follows that there must be a {\it discrete set of modes} $\omega_n$ in the disk. 
This is a starting point or prototype of current theory of discrete global modes in spiral galaxies.
\\

In my opinion, this point is very important. There are several mechanisms that could be 
responsible for the triggering of spiral density waves, e.g., swing amplification or tidal forcing. 
However, only discrete global modes as the  natural eigenfunctions of the disk can provide an 
explanation for the long-lived spiral patterns in galaxies.     
\\

Later, (2) was re-discovered (for some particular model) by Lau, Lin and Mark (1976) 
with no reference to the original paper.
\\

As to Lin and Shu, the words "asymptotic analysis of the WKB type" one can find in Lin, 
Yuan and Shu (1969) for the first time. Let me remind you again, that the Marochnik and Suchkov (1968a) 
preprint has been sent to Lin and Shu by the end of 1968. And finally, Lin, Yuan and Shu (1969) 
and Marochnik and Suchkov (1969a,c) papers appeared simultaneously. However, only  Marochnik and Suchkov (1969a, c) have given the very important condition (2), which is a direct consequence of the WKB approach. 
\\

\section{SOME FACTS OF INTEREST}

DISPERSION RELATION OF SPIRAL DENSITY WAVES IN THE WKB APPROXIMATION 
\\

Lin and Shu (1966) and Lin, Yuan and Shu (1969) gave the density wave dispersion relation that 
took into account the interstellar gas and only  one rotating stellar subsystem with the 
Schwarzschild distribution function of stars. Marochnik and Suchkov (1968a, 1969a) gave the 
general form of the dispersion relation (Eqs. (I.12) and (10), respectively) that took into 
account galactic subsystems (rotating and non-rotating) with an arbitrary stellar distribution 
function and interstellar gas. Both (I.12) and (10) reduce to Lin, Yuan and Shu (1969)'s 
equation (3.2) in the particular case of one rotating stellar subsystem with the Schwarzschild 
distribution function of stars and interstellar gas.  
\\

ENERGY AND ANGULAR MOMENTUM EXCHANGE BETWEEN UNSTABLE SPIRAL DENSITY WAVES AND STELLAR SUBSYSTEMS.
\\

Marochnik and Suchkov (1968c, 1969b) showed that unstable density waves can exchange energy, 
momentum and angular momentum with stars and can provide exchange of energy, 
momentum and angular momentum between galactic subsystems. These are equations (III.47) 
for tightly wound spiral density waves and (III.54) for not tightly wound waves in 
Marochnik and Suchkov (1968c) and appropriate comments (see also equations (10) and (11) in 
Marochnik and Suchkov (1969b)). The most important thing was that this result 
was independent of the nature of instability.
\\

Lin's camp came to this point several years later (see Mark (1976) and references therein). 
In particular, Mark (1976) re-discovered the fact that unstable waves can exchange angular 
momentum with stellar populations. 
\\

I do not want to disparage the brilliant series of Mark's papers of 1971-1976 in any way. 
Mark was the person who investigated this problem in great detail and obtained outstanding 
results that are described also in great detail in our book (Marochnik and Suchkov, 1996).  
All I would like to do is to establish correct chronological milestones. 
\\

ULTRA HARMONIC RESONANCES
\\

We showed the presence of forbidden regions in stellar disks (between resonance overtones) 
where density waves cannot exist (Marochnik and Suchkov (1968b, 1969a). 
Later, in the 1970s this fact was re-discovered in Lin's camp papers on ultra harmonic resonances. 
\\

\section{COMPARISON WITH OBSERVATIONS}

As I already mentioned, both papers of Lin, Yuan and Shu (1969) and Marochnik and Suchkov (1969a) 
have appeared simultaneously in March of 1969. Both papers dealt, in particular, 
with the comparison of their theories with observations. However, in the frame of their WKB approximation, 
Lin, Yuan and Shu (1969) employed an incorrect approach to compare theory with observations. 
It was a direct consequence of Marochnik and Suchkov (1969a) analysis (see below) and led 
to the far-going consequences.  
\\

Here is why Lin, Yuan and Shu (1969) approach was incorrect.
In the linear asymptotic approximation, the dispersion equation of spiral waves (1) is 
the initial relation for quantitative description. The solution of this equation is of the form
\be
 k=k(\Omega_p, \Omega(R), \mu(r), \sigma(R)) {\mbox ,}
\ee
which expresses a relation between the wave number $k$ and the wave frequency $\omega = m\Omega_p$ 
in terms of the Galaxy parameters: the rotation curve $\Omega (R)$, the density distribution $\mu(R)$, 
and the dispersion of star velocities $\sigma(R)$; where $R$ is the (radial) galactocentric distance, 
$\Omega_p$ the spiral pattern rotation velocity; and $m$, the number of spiral arms.
 Lin, Yuan and Shu (1969), in their approach, used a galactic model which is completely determined 
by three parameters $\Omega, \mu$ and $\sigma$. In their model the total mass of the Galaxy is 
'squeezed' into a thin disk. 
They used $\mu$ as the projected surface density of all galactic stars. 
Their model was also characterized by one and the same rotation velocity $\Omega(R)$, 
and one and the same velocity dispersion $\sigma(R)$. 
\\

As a matter of fact, galaxies consist of many subsystems, each of them with its own set of parameters 
$\Omega_i, \mu_i, \sigma_i$. 
Marochnik and Suchkov (1968a,b,c,d;1969a,b) showed that actually not the total Galaxy mass 
(as Lin, Yuan and Shu (1969) used), but only a small portion of it associated with the 
flat subsystem takes part in the wave motion. The characteristics of this subsystem $\Omega_f, \mu_f, \sigma_f$
 strongly differ from the parameters averaged over all subsystems, which are used by Lin, Yuan and Shu (1969). 
\\

As a result, the observed grand design turns out to be associated with quite different types of 
density waves and these waves correspond qualitatively and quantitatively to the other properties of 
the spiral structure (Marochnik et al., 1972). None of this was, unfortunately mentioned in Pasha's history.
\\

Only several years later, Lin and his associates began to consider the so-called "active disk", 
i.e. a subsystem that really contributes to density waves. If my memory serves me right, 
this work began in 1975-1976 but definitely not in1968-1969. 
\\

I have no wish to disparage the excellent work of the 1960s of Kalnajs (1965), Lin and Shu (1964, 1966), 
Lynden-Bell (1962, 1967), Toomre (1964, 1969) and other. This paper is written to correct 
some historical misperceptions. This was due in part to the political situation in the Soviet Union 
which made it frequently impossible for Soviet scientists to discuss face to face their work and 
progress in their chosen field with their Western colleagues.
\\

{\bf Acknowledgments}
\\
 
I am grateful to Yuri Frankel, Vladimir Korchagin, Robert McCutcheon, Elena Mirskaya,
 Yuri Mishurov, Roald Sagdeev, 
Anatoly Suchkov and Boris Vayner for critical comments and advice that helped me to improve the manuscript. 
A special thanks to Vladimir Korchagin who helped me to update the manuscript and Muzafar Maksumov, 
who provided me with originals of some of 1968 preprints.  
Many thanks to Walter Sadowski who gave me important advice. 

{}

\end{document}